\documentclass[twocolumn,preprintnumbers,amsmath,amssymb,superscriptaddress,prl,showpacs]{revtex4}
\usepackage[dvips]{graphicx}% Include figure files
\usepackage{dcolumn}% Align table columns on decimal point
\usepackage{bm}% bold math
\usepackage[dvips]{epsfig}

\begin{document}

\title{Transient Shear Banding in a Simple Yield Stress Fluid}

\author{Thibaut Divoux}
\affiliation{Universit\'e de Lyon, Laboratoire de Physique, Ecole Normale Sup\'erieure de
Lyon, CNRS UMR 5672, 46 All\'ee d'Italie, 69364 Lyon cedex 07, France.}
\author{David Tamarii}
\affiliation{Universit\'e de Lyon, Laboratoire de Physique, Ecole Normale Sup\'erieure de
Lyon, CNRS UMR 5672, 46 All\'ee d'Italie, 69364 Lyon cedex 07, France.}
\author{Catherine Barentin}
\affiliation{Laboratoire de Physique de la Mati\`ere Condens\'ee et Nanostructures, Universit\'e de Lyon, CNRS UMR 5586, 43 Boulevard du 11 Novembre 1918, 69622 Villeurbanne cedex, France.}
\author{S\'ebastien Manneville}
\affiliation{Universit\'e de Lyon, Laboratoire de Physique, Ecole Normale Sup\'erieure de
Lyon, CNRS UMR 5672, 46 All\'ee d'Italie, 69364 Lyon cedex 07, France.}
\date{\today}

\begin{abstract}
We report a large set of experimental data which demonstrates that a simple yield stress fluid, i.e. which does not present aging or thixotropy, exhibits transient shear banding before reaching a steady state characterized by a homogeneous, linear velocity profile. The duration of the transient regime decreases as a power law with the applied shear rate $\dot\gamma$. This power law behavior, observed here in carbopol dispersions, does not depend on the gap width and on the boundary conditions for a given sample preparation. For $\dot\gamma\lesssim 0.1$~s$^{-1}$, heterogeneous flows could be observed for as long as 10$^5$~s. These local dynamics account for the ultraslow stress relaxation observed at low shear rates.
\end{abstract}
\pacs{83.60.La, 83.50.Ax, 83.50.Rp}
\maketitle

Yield stress fluids (YSF) encompass a large amount of everyday-life and industrial complex fluids ranging from hair gels, cosmetic creams, or toothpastes to many food products or even concrete and drilling muds. They respond elastically below a certain stress threshold $\sigma_c$, known as the yield stress, whereas they flow as liquids if stressed above $\sigma_c$. The behavior of YSF still raises many fundamental and practical issues, in particular in the vincinity of this solid-to-fluid transition, often referred to as the unjamming transition \cite{Weeks:2007}. Recently it was proposed to categorize YSF into two groups: (i) those with aging and memory effects that present {\it thixotropic} properties (clay suspensions, colloidal gels, adhesive emulsions) and (ii) those for which such effects can be neglected, which are called {\it simple} YSF (foams, carbopol ``gels'', nonadhesive emulsions) \cite{Ragouilliaux:2007,Moller:2009b}. Since they present glassy-like features, thixotropic materials have received a lot more of attention in the past decade \cite{Barnes:1997,Moller:2006,Coussot:2007}. Their yielding properties result from a competition between aging and shear-induced rejuvenation which leads to a {\it shear banding} phenomenon involving a critical shear rate $\dot \gamma_c$ \cite{Coussot:2002a,Rogers:2008,Moller:2008,Ovarlez:2009}. Indeed, for an imposed shear rate $\dot \gamma>\dot \gamma_c$, the whole fluid flows, while for $\dot \gamma<\dot \gamma_c$, one observes the coexistence of a solid region with a vanishing local shear rate $\dot \gamma_{\rm loc}=0$ and a fluidized region sheared at $\dot \gamma_{\rm loc}=\dot \gamma_c$. Besides, various models in terms of hidden local degrees of freedom such as the fluidity \cite{Picard:2002} or the structural ``$\lambda$ parameter'' \cite{Moller:2006,Coussot:2002a} account for the corresponding steady-state velocity profiles. On the other hand, simple YSF show a {\it continuous} yielding transition in the sense that the local shear rate can take any nonzero value. In that case, the flow may appear as localized only due to stress gradients that result from the strong curvature of wide-gap cylindrical cells: in small-gap geometries, steady-state flow profiles are homogeneous above yielding \cite{Ovarlez:2009,Ovarlez:2008,Coussot:2009,Salmon:2003a,Meeker:2004}. Nonetheless, lately, subtle effects have been reported that somewhat blur the above distinction. First, non-thixotropic nonadhesive emulsions, for which magnetic resonance velocimetry coupled to wide-gap rheology shows simple yielding behavior \cite{Ovarlez:2008}, were found to present shear banding when confined into microchannels \cite{Goyon:2008}. Second, thixotropic Laponite suspensions, that exhibit shear banding in a rough geometry, were shown to slowly evolve towards homogeneous flow in the presence of smooth boundaries \cite{Gibaud:2008}. These findings evidence the influence of confinement and boundary conditions on the yielding scenario and call for a systematic investigation of the spatiotemporal dynamics of YSF.

In this Letter we revisit the case of a simple YSF, namely a carbopol ``gel''. Indeed, in spite of a huge body of rheological work, no thorough temporally and spatially resolved study has ever been conducted on simple YSF. In particular, it is generally assumed that the flow remains homogeneous during its evolution towards steady state. Here shear-rate controlled experiments show that (i)~the transient regime from solidlike to liquidlike behavior involves a shear-banded flow; (ii)~the duration $\tau_f$ of this transient regime decreases as a power law of the imposed shear rate: $\tau_f\propto\dot{\gamma}^{-\alpha}$; and (iii)~the flow always turns out to be homogeneous after this transient shear banding. This fluidization scenario is very robust and the exponent $\alpha$ does not depend on the gap width or on the boundary conditions. Our results not only confirm the idea that one cannot observe steady-state shear banding without thixotropy but also demonstrate that a simple YSF can present shear banding in a transient regime that can last surprisingly as long as $10^5$~s.

\begin{figure}[tb]
\begin{center}
\includegraphics[height=6cm]{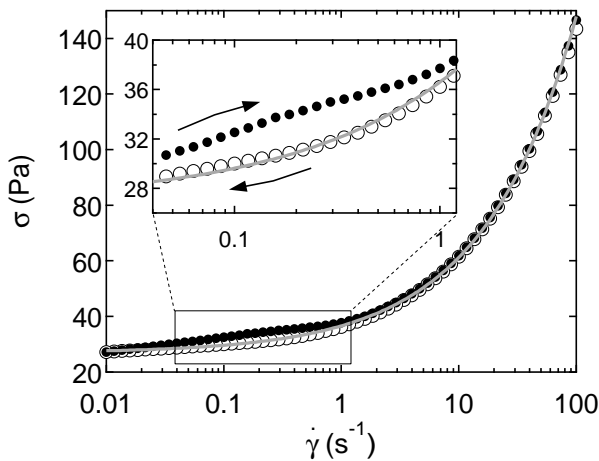}
\end{center}
\caption{\small{The flow curve, shear stress $\sigma$ vs the shear rate $\dot\gamma$, obtained by decreasing $\dot\gamma$ from 100 to 0.01~s$^{-1}$ ($\circ$) and then increasing $\dot\gamma$ from 0.01 to 100~s$^{-1}$ ($\bullet$) with a waiting time of 100 s per point. Note the slight hysteresis for $\dot\gamma\lesssim 1$~s$^{-1}$ which is the signature of a significant slowing down of the system dynamics near the yielding transition (see text). The gray line is the best fit of the decreasing part by the Herschel-Bulkley model: $\sigma=\sigma_c+A\dot \gamma^n$, 
with $\sigma_c=26.9$~Pa, $n=0.55$ and  $A=9.8$~Pa.s$^{-n}$. Inset: zoom over the low shear rate regime. Both walls have a roughness of 60~$\mu$m.}
}
\label{fig.1}
\end{figure}

\begin{figure}[t]
\begin{center}
\includegraphics[height=6.5cm]{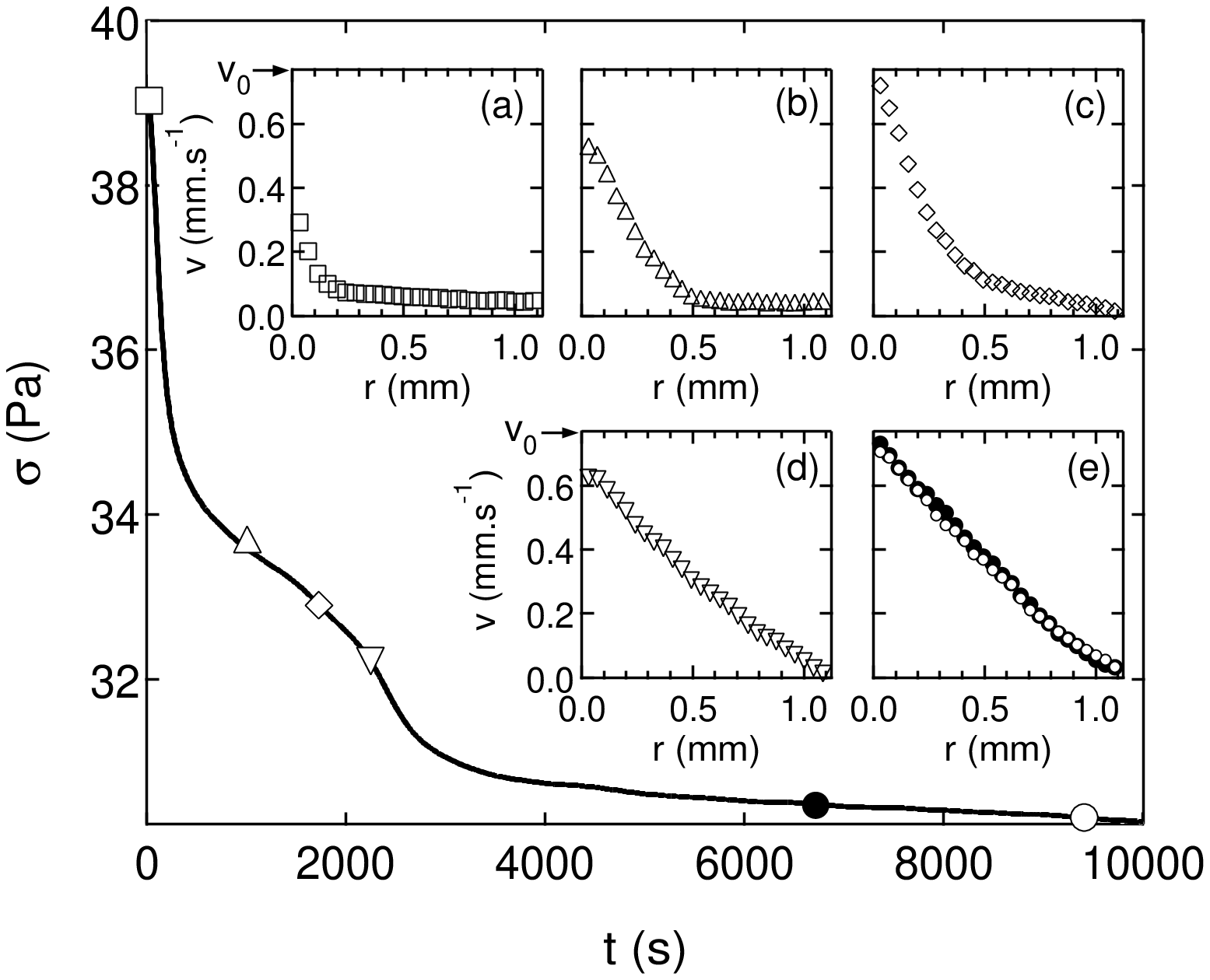}
\end{center}
\caption{\small{The shear-stress $\sigma$ vs time $t$ for a shear rate $\dot\gamma$ = 0.7~s$^{-1}$ applied at $t=0$. Insets: velocity profiles $v(r)$, where $r$ is the distance to the rotor, at  different times [symbol, time (s)]: ($\square$,~27); ($\bigtriangleup $,~1013); ($\diamond $,~1730); ($\bigtriangledown $,~2250); ($\bullet $,~6778); ($\circ $,~9413). The arrow indicates the rotor velocity $v_0$. Both walls have a roughness of 60~$\mu$m. The two-step relaxation of $\sigma(t)$ is strongly correlated to the evolution of $v(r)$ (see text).}
}
\label{fig.2}
\end{figure}

{\it Experimental set-up and protocol -}
Our working system is made of carbopol ETD~2050 dispersed in water at 1\%~wt. This system is referred to as a carbopol ``gel'' in the literature although its microstructure shows a honeycomb-like assembly of swollen polymer regions with typical size 5--20~$\mu$m \cite{Kim:2003,Piau:2007}. Carbopol ``gels'' are known to be non-aging, non-thixotropic simple YSF \cite{Coussot:2009,Moller:2009a}. Here our samples are seeded with micronsized glass spheres \cite{Protocol} in order to use ultrasonic speckle velocimetry (USV) \cite{Manneville:2004} simultaneously to standard rheological measurements with an Anton Paar MCR301 rheometer. The addition of such acoustic contrast agents only slightly stiffens the system (by at most 25\%) and rheological measurements performed without glass spheres show the exact same trends as in Figs.~\ref{fig.1} and \ref{fig.2}. %with stress values that are about 10\% smaller.

The carbopol gel is poured into the gap of a rough Couette cell (rotating inner cylinder radius 23.9~mm, gap width $e=1.1$~mm and height~28 mm). The surface roughness of 60~$\mu$m obtained by gluing sandpaper to both walls was chosen to be of the order of the microstructure size in order to avoid wall slip. To ensure that the strain accumulated during loading into the cell has no influence, the sample is systematically presheared for 1 min at +1000~s$^{-1}$ and for 1 min at -1000~s$^{-1}$. We then check that a reproducible initial state is reached by measuring the viscoelastic moduli $G'=100\pm 10$~Pa and $G''=12\pm 2$~Pa for 2~min. Finally the sample is left at rest for 1 min before starting the experiment.

{\it Constitutive behavior -}
Figure~\ref{fig.1} shows that the Herschel-Bulkley model accounts perfectly for the flow curve of our carbopol system in good agreement with previous works \cite{Coussot:2009,Kim:2003,Piau:2007,Moller:2009a}. In particular, the absence of a stress plateau in the flow curve at low shear rates in a rough geometry is usually taken as the signature of simple yielding behavior \cite{Moller:2009b,Meeker:2004}. However we observe a slight hysteresis between decreasing and increasing shear-rate sweeps for $\dot\gamma \lesssim 1$~s$^{-1}$ (see inset of Fig.~\ref{fig.1}). Such an effect is also noticeable but not discussed in previous reports \cite{Coussot:2009,Moller:2009a}. We checked that the size of the hysteresis cycle decreases as the waiting time is increased from 100~s to 500~s per point. We shall see below that it may take a surprisingly long time to reach steady state at low shear rates. Therefore we will attribute the slight hysteresis in the flow curve to a critical slowing down of the system dynamics close to yielding and not to any memory effect.

\begin{figure}[tb]
\begin{center}
\includegraphics[height=6cm]{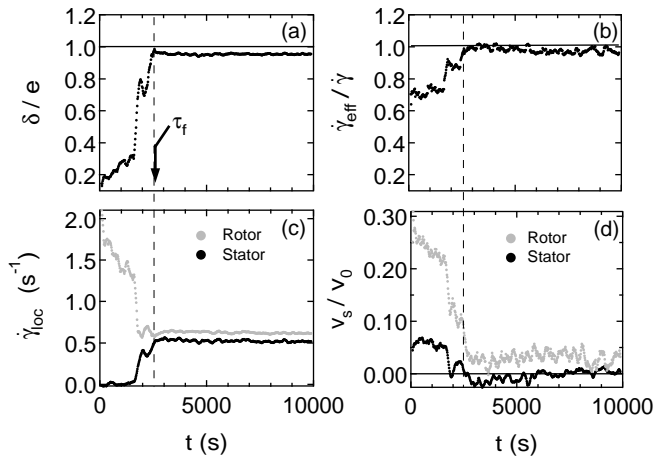}
\end{center}
\caption{\small{Analysis of the velocity profiles recorded every 18~s during the experiment at $\dot\gamma =0.7$~s$^{-1}$ shown in Fig.~\ref{fig.2}. (a)~Width $\delta$ of the fluidized shear band normalized by the gap width $e$ vs time. The shear band invades the whole gap over a characteristic time $\tau_f$. (b)~Evolution of the effective shear rate $\dot\gamma_{\rm eff}$ normalized by the imposed shear rate $\dot\gamma$ vs time. (c)~Local shear rates $\dot\gamma_{\rm loc}$ close to the walls vs time. (d)~Slip velocities $v_s$ normalized by the rotor velocity $v_0$ vs time. In (c) and (d), gray (black resp.) symbols correspond to the rotor (stator resp.). The errorbars indicate the uncertainty due to our spatial resolution of 40~$\mu$m and to the procedure used to fit the velocity profiles \cite{Analysis}.}
}
\label{fig.3}
\end{figure}

{\it Start-up experiments -}
The above sweep test shows that a better method to probe yielding is to impose a constant shear rate $\dot\gamma$ over a long duration ($>500$~s) and to record the shear stress $\sigma(t)$. Meanwhile we measure velocity profiles at about 15 mm~from the cell bottom through USV \cite{Manneville:2004}. Such a combined technique gives access to both global and local dynamics of the fluidization process. As shown in Fig.~\ref{fig.2} for $\dot\gamma=0.7$~s$^{-1}$, the shear stress follows a two-step relaxation which can be analyzed in light of the velocity profiles: (i) for $t\lesssim 2000$~s, a shear band is present at the rotor side and slowly expands across the gap [Fig.~\ref{fig.2}(a)--(c)] and (ii) for $t\gtrsim 2000$~s, the velocity profiles becomes linear within a few 100~s [Fig.~\ref{fig.2}(d)] and the flow remains homogeneous and stationary up to experimental uncertainty for the rest of the experiment [Fig.~\ref{fig.2}(e)]. The ``fast'' fluidization at $t\simeq 2000$~s is better seen in Fig.~\ref{fig.3}(a) where the width $\delta$ of the shear band is plotted against time. Comparing Fig.~\ref{fig.2} and Fig.~\ref{fig.3}(a), we interpret the plummet of $\sigma(t)$ around $t=2000$~s as the rheological signature of the end of shear-banding.

The velocity profiles of Fig.~\ref{fig.2}(a),(b) clearly point to the presence of strong wall slip at the rotor even though rough surfaces were used. The complete data analysis \cite{Analysis} presented in Fig.~\ref{fig.3}(b)--(d) shows that smaller but noticeable slip is also present at the stator and that slip velocities as well as the local shear rate close to the rotor decrease concomittantly to the growth of the fluidized band [Fig.~\ref{fig.3}(c),(d)]. Wall slip becomes completely negligible %, i.e. the effective shear rate $\dot\gamma_{\rm eff}$ experienced by the material coincides with the applied shear rate $\dot\gamma$,
only when homogeneous flow is reached [Fig.~\ref{fig.3}(b)].

The above experiments and analysis were repeated for shear rates ranging from 0.01 to 20~s$^{-1}$ and for various gap widths $e=1.1$--3~mm and wall roughnesses (``smooth'' raw Plexiglas with a roughness of about 15~nm or 60~$\mu$m rough sandpaper). The scenario described in Figs.~\ref{fig.2} and \ref{fig.3} was found to be very robust. Note however that, depending on the applied shear rate, the bump in $\sigma(t)$ does not always show as clearly as in Fig.~\ref{fig.2}: for $\dot\gamma\gtrsim$~2~s$^{-1}$, it gets buried into the initial decay while for $\dot\gamma\lesssim$~0.2~s$^{-1}$ the stress jump becomes very small. Therefore local velocity measurements are essential to assess the nature of the stress relaxation towards steady state.

\begin{figure}[tb]
\begin{center}
\includegraphics[height=10.5cm]{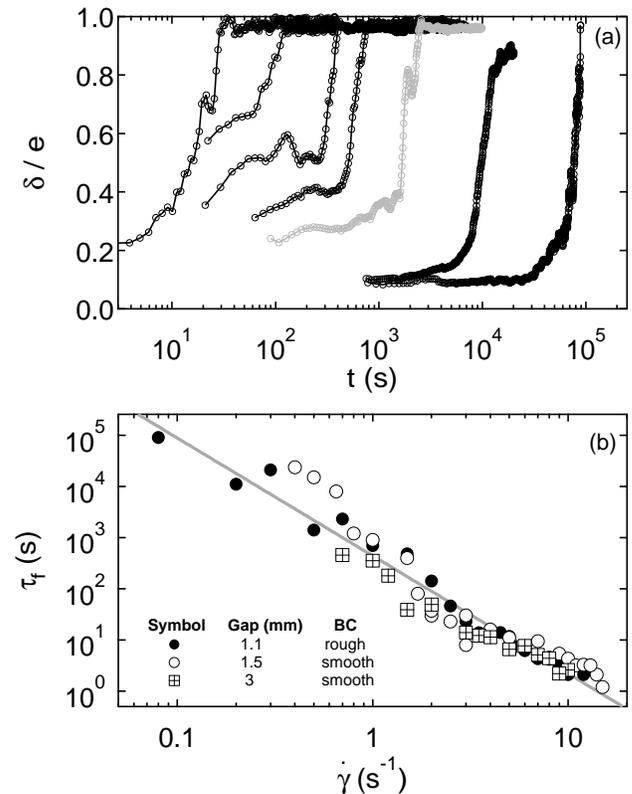}
\end{center}
\caption{\small{(a) Width $\delta$ of the shear band normalized by the gap width $e$ vs time
for different applied shear rates $\dot\gamma =3.0, 2.0, 1.5, 1.0, 0.7 {\rm~(gray)}, 0.2, 0.08$~s$^{-1}$ from left to right. Both walls have a roughness of 60~$\mu$m. The time interval between two successive velocity profiles was set to be at least 1~s and inversely proportional to $\dot\gamma$ in order to focus on the long-time relaxation. For each curve, the fluidization time $\tau_f$ is extracted as indicated in Fig.~\ref{fig.3}(a). (b)~Fluidization time $\tau_f$ vs the applied shear rate $\dot\gamma$ for various gap widths and boundary conditions (BC). The gray line is a power-law fit $\tau_f = \beta /\dot \gamma ^{\alpha}$ with $\alpha = 2.3\pm 0.1$ and $\beta=272\pm 11$~s$^{1-\alpha}$.
%for the 1.1~mm gap with walls of roughness 60~$\mu$m ($\bullet$), $\alpha = 3.1\pm 0.1$ for a 1.6~mm gap with a rough (60~$\mu$m) rotor and a smooth stator ($\square$) and $\alpha = 2.1\pm 0.1$ for a $3$~mm gap with smooth walls ($\bigtriangleup $).
} 
}
\label{fig.4}
\end{figure}

In all cases a shear band is observed transiently over a ``fluidization time'' $\tau_f$ defined as the time when homogeneous flow is reached as indicated in Fig.~\ref{fig.3}(a). The width of the fluidized zone keeps a similar evolution as $\dot\gamma$ is varied, although the initial value of $\delta$ seems to depend on $\dot\gamma$ [Fig.~\ref{fig.4}(a)]. As a major result of this Letter, we observe that $\tau_f$ decreases as a power law for increasing shear rates [Fig.~\ref{fig.4}(b)]. For $\dot\gamma \lesssim 1$~s$^{-1}$, $\tau_f$ is larger than 100~s and fluidization times as long as $10^5$~s were measured for $\dot\gamma \lesssim 0.1$~s$^{-1}$. Such ultraslow dynamics at low shear rates undoubtedly account for the small hysteresis in the flow curve of Fig.~\ref{fig.1} as well as for the stress relaxations reported in Fig.~\ref{fig.2} and in \cite{Coussot:2009}. The absence of any clear dependence on the gap width as well as the fact that flow curves and stress relaxations very close to those of Figs.~\ref{fig.1} and \ref{fig.2} were measured in both cone-and-plate and plate-plate geometries (data not shown) imply that the fluidization process does not significantly depend on geometry-induced stress and strain heterogeneities.

{\it Discussion -} As already implied by recent works \cite{Coussot:2009,Moller:2009a}, we confirm through local velocity measurements in small-gap Couette cells that steady-state velocity profiles are homogeneous for a simple YSF. More importantly our study brings about a number of findings that contradict the simple picture of a fast, continuous transition from solidlike to liquidlike behavior under a constant shear rate. A central result is the scaling law $\tau_f\propto\dot\gamma^{-\alpha}$ for the fluidization time. From the experimental point of view, it means that it is crucial to check whether the flow has indeed reached a steady-state, especially at low shear rates, before drawing conclusions from standard rheological data such as the flow curve or the stress response. Note that the stress in Fig.~\ref{fig.2} has not fully relaxed after $10^4$~s even though the velocity profiles measured roughly at middle height of the Couette cell are perfectly stationary. This suggests that fluidization may not occur in a fully homogeneous way along the vorticity direction due to end effects at the cell bottom or at the free surface. %Only two-dimensional flow imaging could clarify this issue.

The above fluidization is reminiscent of the slow fragmentation and erosion process triggered by slippery walls in thixotropic Laponite suspensions \cite{Gibaud:2008}, for which the characteristic time was also shown to scale as a power law with a similar exponent but with a critical shear rate $\dot\gamma^*$ below which fluidization is never observed: $\tau_f\propto(\dot\gamma-\dot\gamma^*)^{-2.1}$ \cite{Gibaud:2009}. Together with the recent simulation results on sheared attractive particles showing transient shear banding \cite{Chaudhuri:2010} and earlier observations of ultraslow transients in concentrated emulsions \cite{Becu:2005}, the present findings suggest that a generic yielding mechanism involving fluidization times with a critical-like behavior may be at play and go beyond the distinction between simple and thixotropic YSF based on the steady-state.

Providing a model for such a mechanism, in particular for the exponent $\alpha$, appears as a major challenge for future work. Although $\alpha$ is very robust for a given batch as shown in Fig.~\ref{fig.4}, it should be mentioned that values of $\alpha\simeq 3$ were obtained for a different sample preparation, hinting that $\alpha$ is a material parameter that may vary from one simple YSF to the other. Up to now theoretical approaches have focused on steady states \cite{Coussot:2002a,Picard:2002,Bocquet:2009} or on the viscosity bifurcation under imposed shear stress \cite{Moller:2008,Fielding:2009} so that a comparison of our experiments with existing models is not yet available. Still the most striking feature of the transient regime is the presence of shear banded flows with strong wall slip in spite of rough boundary conditions that are usually assumed to prevent slip phenomena. This observation suggests a yielding mechanism where the initially solidlike material suddenly fails at the inner wall independently of boundary conditions and leaves a strongly sheared lubrication layer at the rotor which then fluidizes the solidlike sample. Moreover the sudden acceleration of $\delta(t)$ around $\tau_f$ [Fig.~\ref{fig.4}(a)] and the absence of gap dependence rule out propagative or diffusive processes but rather points to an avalanche-like behavior although no thixotropy is present. Such ingredients should be included in future tentative models for the observed dynamics.

% A specific investigation at early times focusing on the shear band nucleation will be conducted to confirm such a yielding scenario that would lead to transient shear banding without aging or thixotropic effects.

\begin{acknowledgments}
We thank the Anton Paar company for lending us an MCR301 rheometer, Y. Forterre for providing the carbopol and E.~Bertin, L.~Bocquet, A.~Colin, P.~Chaudhuri and N.~Taberlet for fruitful discussions.
\end{acknowledgments}


\begin{thebibliography}{}
\bibitem{Protocol} The detailed protocol is as follows: we first add 0.5\%~wt of hollow glass spheres (Potters, Sphericel, mean diameter 6~$\mu$m) in ultrapure water; pH increases roughly from 7 to 8. As carbopol is hydrosoluble only for pH$<7$, we add one or two drops of concentrated acid to make the pH drop to 6. The mixture is heated at $50^{\circ}$C and the carbopol powder is carefully dispersed at 1\%~wt under magnetic stirring at 300~rpm for 40~min then left at rest at room temperature for another 30~min, after which pH~$\simeq 3$. Finally we neutralize the solution with NaOH until pH~$=7$ while stirring manually. This leads to a carbopol ``gel'' which is finally centrifuged (10~min at 2500~rpm) to get rid of trapped bubbles. %This protocol allows us to obtain a very reproducible carbopol system seeded with small glass spheres.
\bibitem{Analysis} Linear fits of the velocity profiles over about 150~$\mu$m are used to estimate the local shear rates $\dot\gamma_{\rm loc}$ close to each wall and extrapolate the sample velocities $v(0)$ and $v(e)$. The effective shear rate is defined as $\dot\gamma_{\rm eff}=[v(0)-v(e)]/e$. The width of the shear band is obtained as the abscissa of the intersection between the two fits. The slip velocity at the rotor (stator resp.) is given by $v_s=v_0-v(0)$ ($v_s=v(e)$ resp.) where $v_0$ is the rotor velocity.
\end{thebibliography}
\end{document}